\documentclass[12pt, a4paper]{article}
\usepackage[top=2.54cm, bottom=2.54cm, left=2.54cm, right=2.54cm]{geometry}
\usepackage[T1]{fontenc}
\usepackage[utf8] {inputenc}
\usepackage[german,english]{babel}
\usepackage{array}
\usepackage{graphicx}
\usepackage[longnamesfirst]{natbib}
\usepackage{setspace}
\usepackage{rotating}
\usepackage{amsmath}
\usepackage{amssymb}
\usepackage{amsfonts}
\usepackage{amsthm}
\usepackage{rotating}
\usepackage{pdflscape}
\usepackage{afterpage}
\usepackage{enumitem}
\setlist{nolistsep}
\usepackage{tikz}
\usetikzlibrary{positioning}
\usetikzlibrary{arrows,decorations.markings}
\usetikzlibrary{patterns,snakes}
\usetikzlibrary{graphs}
\usetikzlibrary{calc, shapes, backgrounds, fit}

\usepackage{verbatim}
\usepackage{booktabs}
\usepackage{subcaption}
\usepackage{caption}

\usepackage{hyperref}
\hypersetup{
	pdfstartview = FitH,
	pdfauthor = {...},
	pdftitle = {...},
	pdfkeywords = {...; ...; ...; ...},
	colorlinks = true,
	linkcolor = blue,
	urlcolor = blue,
	citecolor = blue,
	linktocpage=true
}

\begin{document}

\title{\textsc{\LARGE{The role of unobservable characteristics in friendship network formation}}}

\author{Pablo Bra\~nas-Garza\thanks{Universidad de Loyola Andalucía. E-mail: \texttt{branasgarza@gmail.com}} \and Lorenzo Ductor \thanks{Universidad de Granada. E-mail: \texttt{lductor@ugr.es}}
\and Jarom\'ir Kov\'a\v{r}\'ik \thanks{Universidad del Pa\'is Vasco UPV/EHU and University of West Bohemia. Email: \texttt{jaromir.kovarik@ehu.eus} \newline We acknowledge the financial support of MINECO-FEDER (PGC2018-093506-B-I00, PID2019-106146GB-I00 and PID2019-108718GB-I00), Excelencia—Andalucía (PY18-FR-0007), the Basque government (IT1336-19), GAČR (17-25222S) and Fundaci\'on Ram\'on Areces.}}

\date{\today}

\maketitle
\thispagestyle{empty}

\begin{abstract}

Inbreeding homophily is a prevalent feature of human social networks with important
individual and group-level social, economic, and health consequences. The
literature has proposed an overwhelming number of dimensions along which human
relationships might sort, without proposing a unified empirically-grounded
framework for their categorization. We exploit rich data on a sample of University
freshmen with very similar characteristic - age, race and education- and contrast the relative importance of observable vs. unobservables
characteristics in their friendship formation. We employ Bayesian Model Averaging,
a methodology explicitly designed to target model uncertainty and to assess
the robustness of each candidate attribute while predicting friendships. We show
that, while observable features such as assignment of students to sections, gender,
and smoking are robust key determinants of whether two individuals befriend each
other, unobservable attributes, such as personality, cognitive abilities, economic
preferences, or socio-economic aspects, are largely sensible to the model specification,
and are not important predictors of friendships.

\end{abstract}

\newpage
\setcounter{page}{1}

\section{Introduction}\label{sec: intro}


Individuals tend to interact and form relationships with others who are more similar to one's self than randomly chosen members of the population. This phenomenon, known as homophily or “love of the same” \citep{lazarsfeld1954friendship}, is one of the most prevalent and robust features of human social networks and has been widely studied across fields \citep{lazarsfeld1954friendship, mcpherson2001birds, moody2001race, kossinets2009origins}. The ubiquitous presence of homophily directly translates into human interactions segregated across a variety of dimensions. Since the resulting segregation patterns have important consequences on behavior, diffusion, inequality, and social mobility \citep{glaeser2000social, ruef2003structure, centola2011experimental, bramoulle2012homophily, jackson2017economic, zeltzer2019gender}, it is of utmost importance to understand the origins, nature, and consequences of homophily.

Rather than analyzing the origins or consequences, the objective of this article is to test the relative importance of the large set of differing dimensions of homophily in friendship networks analyzed in previous studies (see \cite{mcpherson2001birds} for an influential review of the literature). Abstracting from whether the studies document induced or choice homophily or rather social influence \cite{kossinets2009origins, currarini2010identifying}, there exists large and robust evidence that friendships sort on gender, age, race, ethnicity, religion, and socio-economic status \citep{wittek2020fighting, kruse2016neighbors, ibarra1992homophily, moody2001race, marmaros2006friendships, kossinets2009origins, currarini2010identifying, zeltzer2019gender}. In the same vein, meeting constraints--proxied by living in the same neighbourhoods, belonging to the same club, working in the same firm, studying the same major, living in the same dorm, and so on--are important predictors of who interacts with whom \citep{glaeser2000social, marmaros2006friendships, kossinets2009origins, mayer2008old}. Some studies have focused on more subtle but still detectable dimensions, such as smoking \cite{christakis2008collective} and obesity \citep{christakis2007spread, valente2009adolescent, de2010obesity}; others have analyzed factors imperceptible by a naked eye, such as political orientation, academic performance, cognitive abilities, behavioral or personality traits among others \citep{almack1922influence, marmaros2006friendships, conti2013popularity, mayer2008old}.\footnote{The number of studies proposing one or multiple dimensions of homophily in social networks is large and reviewing all of them is out of scope of this article.}

Although the existing work has established the baseline for homophily theories, the literature has proposed an overwhelming number of dimensions along which human relationships might sort, without providing a unified framework according to which we can categorize the different dimensions to determine their relative importance.\footnote{\cite{lazarsfeld1954friendship} distinguish between status homophily and value homophily, but this categorization is based on the nature of the factor rather than on empirical grounds.} That is, there is no systematic analysis assessing how important and robust the proposed predictors of friendship are from a statistical perspective. 

The objective of this paper is to provide a step toward in characterizing the potential dimensions of homophily based on their empirical relevance. To that aim, we point out one problem of the existing literature: the typical approach is to analyze a \textit{small} and/or \textit{exogenously given} set of potential dimensions of homophily, depending on the data and variables at researchers' disposal. Nevertheless, these approaches can lead to pre-testing bias and increase the ”researcher's degrees of freedom” and ”p-hacking” if the estimation is conducted using multiple combinations of regressors with the aim of obtaining statistically significant estimates \citep{simmons2011}. In addition, many human characteristics are highly correlated in real life: gender is frequently correlated with self-confidence, prosocial, and risk attitudes \citep[see e.g.][]{barber2001boys, croson2009gender}, while race predicts a large array of socio-economic outcomes \citep{guterbock1983race, altonji1999race}. 

These correlations in turn generate two classic statistical issues while estimating homophily using standard approaches. Focusing on a small number of characteristics increases the probability of accepting a particular determinant of friendship formation incorrectly (see below for concrete examples) due to the omitted-variable problem,  which leads to biased--and even inconsistent--estimates.

A converse issue arises if several correlated factors are included into one model. Such a multicollinearity issue leads to a well known problem regarding the interpretation of the individual estimates and cast doubts on the (non-)redundancy of the individual factors. This multicollinearity problem may result in a loss of efficiency, leading to an incorrect rejection of a determinant. Thus, both statistical issues can lead to serious misunderstanding of origins and consequences of homophily.

Using the statistical jargon, the overwhelming variety of potential factors that might influence network formation generates a problem known as \textit{model uncertainty}, arising from the lack of a unified framework for examining the drivers of friendship. One natural way to think about model uncertainty is to admit that we do not know which is the “true” model and attach probabilities to different models \citep{sala2004}. This is the main idea behind Bayesian Model Averaging (BMA), which consists in estimating all possible combinations of the regressors, comparing the performance of all the models resulting from the different combinations, and taking a weighted average over all the candidate models, where the weights are the probabilities that the candidate model is the true model \citep{hoeting1999bayesian}. In our analysis, we account for uncertainty in the friendship formation model and identify robust determinants of link formation in friendship networks by using a BMA panel-data approach.\footnote{For the sake of robustness, we also employ the Weighted Least Squares estimation method to identify the determinants. Our conclusions are robust to the modeling approach; see Section \ref{sec: rob}.} To the best of our knowledge, this is the first study that addresses model uncertainty while identifying the main drivers of friendship. 

In this paper, we study potential determinants of friendship formation among the freshmen of a faculty at a major Spanish University ($n=273$). The data at our disposal have several particular features. First, we mapped the friendship networks twice, once at the first week of their freshmen year and once at the very end of the same academic year. Such timing provides a unique opportunity to examine the determinants of the emergence of friendships in the group. Although some links already exist in the beginning, the first-week network resembles a scarcely connected randomly generated network. Most of the elicited friendships were established between the two periods and we can control for preexisting relationships using the first network elicitation. 

Second, we collected a rich set of individual characteristics for each student, enabling to examine the role of homophily on large array of dimensions. We broadly classify the dimensions under study as observable and unobservable (or imperceptible) and contrast their importance in friendship formation. We define an attribute as observable if it can be detected by mere observation. Examples of such characteristics are race, gender, geographical/physical proximity, obesity, or whether one smokes. We classify as unobservable or imperceptible characteristics those attributes that require more intimate disclosure or a deeper conversation to be detected; these include family background, cognitive and personality traits, or economic preferences. 

As for our starting hypothesis, on the one hand, there is no reason to believe \textit{a priori} that the observable characteristics should play a more important role that the unobservable ones, or viceversa. On the other hand, one may argue that observable factors are more salient and, hence, the primary drivers of friendship formation, while homophily on unobservable attributes may only arise after certain amount of personal interaction. In line with the latter, \cite{mcpherson2001birds} note that --in contrast to, say, gender or race-- socio-economic indicators are more sensible on the strength of the relationships. \cite{mayer2008old} estimate the determinants of friendships using an array of race and socio-economic indicators both separately and jointly. If separately considered, most indicators predict friendships. However, once all indicators are included into one regression, the influence of the socio-economic attributes weakens and some become insignificant. Similarly, \cite{girard2015} report that, although their behavioral indicators all correlate significantly with friendships, the magnitude, significance, and even the sign of their estimated impact on friendships in multivariate regressions are sensible to the set of dependent variables. These issues are akin to the omitted-variable and collinearity issues discussed above. Therefore, we prefer not to be inclined toward any of the two--or any other--alternative hypothesis and let the data speak. The methodology that we employ is explicitly designed to target such \textit{ex-ante} ambiguity in the underlying hypothesis.
We note that, in contrast to other studies, our sample is largely homogeneous in term of race, ethnicity, age, ``profession", cohort, major, and religion, all belonging among the classic determinants of homophily \cite{mcpherson2001birds}. We belief that this is an advantage of our study: such homogeneity in terms of observable features maximizes the chances that our subjects organize along the unobservable, imperceptible traits under study.

We briefly summarize our findings in what follows. Applying the BMA uncovers that the most important determinants of friendship formation are common gender, common study section (reflecting the importance of geographical location), and smoking habits.
In a stark contrast, none of the unobservable factors capturing individual preferences, socio-economic indicators, and individual personality and behavioral traits play any robust role in friendship formation in our data. Since women and men form networks differently \cite{ductor2018gender}, we show that friendship formation differs between women and men. We found that common gender, smoking habits and common section are specially important to explain the friendships of women. However, the only robust determinant for men is common class section among our candidate factors.

The rest of the paper proceeds as follows. Section \ref{sec:methodology} describes the data and lays out the empirical strategy. Section \ref{sec: findings} presents our findings. Section \ref{sec: rob} assesses the robustness of the results. We discuss the implications of our findings in Section \ref{sec: discussion}.


\section{Materials and Methods}
\label{sec:methodology}

We exploit a data set collected from all freshmen ($n=273$) in the Faculty of Economic and Business Sciences  at the Universidad de Granada, Spain, over the whole academic year 2010-2011. These data contain rich information on students' individual characteristics and their social networks.\footnote{The data were previously employed in an analysis of the role of prenatal exposure to sex hormones and social integration \citep{kovavrik2017} and altruism \citep{branas2013second}, respectively.} The data elicitation was approved by the Ethical Committee of the Universidad de Granada and all subjects provided informed written consent. Anonymity was also assured according to the Spanish Law 15/1999 for Personal Data Protection.

We use this database because it exhibits three particular features providing a unique opportunity to examine the role of both perceptible and imperceptible determinants of friendship formation, the main question of this study.

\subsection{Sample and section distribution}

The first interesting feature of the data is the homogeneity of the sample in terms of race, ethnicity, age, religion, major, cohort, and ``profession'' (Table \ref{Table indvchar}). By definition, all participants are first-year students (without any profession), studying Economics. In addition, our subjects are mostly Caucasian students (99\% of them), aged 17-18 (80\% of them), and catholic  (61\% of them).

\begin{table}[h!]
\centering
\caption{Participant's characteristics}
\begin{tabular}{lcc} \hline
 &  (1)  & (2) \\
Variable  & Observations & Frequency\\\hline
\textbf{Year of birth} &  &  \\\hline
$<$1990 & 27 & 14\% \\
1990  & 11 & 6\% \\
1991  & 43 & 22\% \\
1992  & 113 & 58\% \\\hline

\textbf{Religion} &  &  \\\hline
Catholic & 120 & 62\% \\
Non-Believer/Agnostic & 62 & 32\%\\
Other & 9 & 6\% \\\hline

\textbf{Gender} &  &  \\\hline
Male & 136 & 56\% \\
Female & 108 & 44\% \\\hline

\textbf{Ethnicity} &  &  \\\hline
Caucasian  & 202 &  99\%\\
Other  & 3 & 1\%\\\hline

\textbf{Study/class sections} &  &  \\\hline
A  & 61 & 27\% \\
B  & 64 & 29\%\\
C  & 48 & 21\%\\
D  & 51 & 23\%\\\hline

\textbf{Household Monthly Income} &  &  \\\hline
Income $=<$ 1500 & 69 & 36\% \\
1500 $<$Income $=<$ 2500 & 63 & 32\% \\
2500 $<$Income $=<$ 3500 &  35 & 18\% \\
Income $>$ 3500 & 27 & 14\%  \\\hline
\end{tabular}
\begin{minipage}{16cm}
   \footnotesize 
  \end{minipage}
\label{Table indvchar}
\end{table}

Therefore, out of the classic determinants of friendships and homophily \citep{mcpherson2001birds}, we only expect gender, proximity (see below), and similar socio-economic background to affect who befriends whom. We believe that, due to such homogeneity, there is a large chance that friendships sort out by the more imperceptible, unobservable characteristics under scrutiny (see Section \ref{sec: var} and Table \ref{Table varlist}).

At beginning of their first year, the School distributes the incoming students randomly into study/class sections\footnote{This is the standard procedure and unrelated to our study.}. In the case of the 2010-2011 cohort, the students were randomly assigned into four sections (labelled as A, B, C, and D). No student has any ability to self-select into a particular section and they are explicitly advised to attend the classes of their corresponding section. Following the previous literature \citep[e.g.][]{marmaros2006friendships, carrell2013natural}, we use the section distribution as a measure of location or geographical distance, determining meeting opportunities of students. The random assignment provides a causal interpretation regarding the role of meeting opportunity in link formation \citep{marmaros2002peer}.

\subsection{Networks}

The second advantage of the dataset stems from the fact that networks were elicited twice, during the first week of the academic year in October 2010 (Period 1) and at the end of the academic year in May 2011 (Period 2). All students were initially invited to disclose the names of their friends at the \textit{beginning} of the \textit{first} year. This is an important aspect of the data because it enables us to control for any pre-existing relationships and focus on the social dynamics taking place within the class during the first academic year. The students were again surveyed at the end of the school year when one may expect to observe a standard social network.\footnote{See \cite{kovavrik2017} for more information regarding the network elicitation.}

For the analysis, we mapped the elicited friendships in two undirected networks using subjects as nodes and friendship nominations as links under the conditions that a link between two individuals exists if both named each other as a friend. That is, we focus on the network of reciprocal relations. Since link reciprocity is one of Granovetter's \citep{granovetter1973strength} conditions for a relationship to be considered strong, we focus on reciprocal ties to make sure that we analyze friendship rather than acquaintances.

As expected, the network in period 1 was sparsely connected, suggesting that students had very few ties with their peers at the beginning of the school year. In stark contrast, the end-of-the-year network resembles classic socially generated networks in most respects (e.g. skewed degree distribution, high reciprocity of connections, network transitivity, positive assortativity etc.; see Table 1 and Figure 3 in \cite{kovavrik2017}). This time setting and the fact that the friendship network was almost empty in period 1 allow us to evaluate the causal role of observable characteristics (such as gender or smoking behaviour) and imperceptible features (e.g., personality or economic preferences) in the link formation process of friendship networks.\footnote{The fact that the initial network is almost empty allows us to consider simple probability models in the estimation of the formation of ties.}

Since participation was voluntary, some subjects were not active in both periods (similar consideration applies in case of the data introduced in Section \ref{sec: var}). The number of students in period 1 and 2 was 204 and 246, respectively; 168 students participated in both periods. In our empirical analysis, we focus on the 168 subjects who participated in both periods, since we can estimate their link formation in Period 2 conditional on the prior network in Period 1 or focus on people who have not had any link in Period 1 (our robustness check in Section \ref{sec: rob}).\footnote{We are not able to identify friendships made between period 1 and 2 that were broken before period 2.}

We use the 168 subjects to create all possible pairs of potential friendships among the 168 subjects, removing duplicated pairs. Thus, we have a pairwise panel data with 8515 observations. The dependent variable of interest is friendship indicating whether two individuals $i$ and $j$ nominated each other in Period 2 as friends. This variable $g_{ij,2}=1$ if both $i$ and $j$ did; $g_{ij,2}=0$ otherwise.\footnote{To illustrate the definition of pairwise panel and our dependent variable, consider the following example. We have a two-column matrix, where column 1 contains the id of individual $i$ and column 2 the id of individual $j$. For individual $i=1$, we create $n-1=167$ observations to indicate whether person 1 is a friend of each $j=2,3,..,168$. The dependent variable, $g_{1,3}=1$ if individual individuals 1 and 3 mutually nominate each other as a friend. In addition, the observation $g_{3,1}$ is removed from the sample, since the friendship relationship between 1 and 3 is already accounted for in the observation for $i=1$ and $j=3$.} We then consider the (dis)similarity in observable and imperceptible characteristics between $i$ and $j$ as the determinants of their friendship formation. 

\subsection{Non-network variables}
\label{sec: var}

The third feature of the data worth mentioning is the richness of the available information on the subjects. Apart from the network elicitation, subjects participated in a series of questionnaires, surveys, and incentivized economic experiments designed to collect data on their heterogeneity over their whole freshmen year. In this study, we consider 20 characteristics that might affect the formation of students' friendships. We broadly classify these characteristics as observable (O) and unobservable or imperceptible characteristics (U) and contrast their importance in explaining friendship formation. Observable characteristics are attributes that can be easily observed without any deep interaction, whereas we classified as unobservable or imperceptible those attributes that would require more intimate disclosure or a deeper conversation to be detected. To assess the (dis)similarity of two individuals, we either employ the absolute difference between their values for the characteristic in question or an indicator dummy if they share a characteristic. Table \ref{Table varlist} lists all the characteristics and indicates whether homophily is assessed in our regressions using an absolute value (Transformation = A) or an indicator dummy (Transformation = D). In what follows, we present all the  variables in Table \ref{Table varlist} one by one:
\begin{itemize}
    \item \textbf{Observable characteristics:} 
\begin{enumerate}  \setlength{\itemsep}{0.5em}
    \item \textit{Common gender.} Large literature has documented that age, gender, religion, race, and ethnicity are key determinants of friendship \citep{patacchini2016racial, mcpherson2001birds, currarini2009economic, kossinets2009origins, currarini2010identifying}. Since our sample is homogeneous in terms of age, religion, race, and ethnicity, we only study common gender among these classic observable factors.
   \item \textit{Common section.} Geographical location and spatial proximity are crucial for interaction (\cite{preciado2012does}). For instance, \cite{glaeser2000social} find that students in university dorms are more likely to interact with their roommates and dormmates even if they are assigned to dorms and rooms randomly, \cite{currarini2010identifying} document that friendship formation is largely influenced by meeting chances, and \cite{preciado2012does} show that the log-odds of friendship existence decrease with the geographical distance. In our setting, all the first-year students were randomly distributed into four sections at the beginning of the freshman year. We thus expect homophily in section assignments. We capture such ``location'' using a dummy variable \textit{common section}, which is equal to one if both $i$ and $j$ belong to the same section. 

\item \textit{Body mass index (BMI).} \cite{christakis2007spread} document homophily in BMI as a driver of the epidemic obesity in the U.S. \cite{de2010obesity} show that male friends tended to be similar in their consumption of high-calorie foods. We thus include the absolute difference in BMI of $i$ and $j$ as a potential determinant of them becoming friends. 

\item \textit{Smoking.} Individuals who smoke have been documented to cluster together \citep{christakis2008collective, hsieh2018}. We therefore consider a \textit{smoking} dummy equal to one if both $i$ and $j$ self-report being smokers.

\item \textit{Pre-existing friendship.} Finally, our regressions control for whether $i$ and $j$ had been reported each othe as a friend at the first week of their freshmen year, using a variable \textit{Friends Period 1} that indicates whether $i$ and $j$ named each other in Period 1.

\end{enumerate}
\vspace{0.3em}
\item \textbf{Unobservable characteristics:} 
\begin{enumerate} \setlength{\itemsep}{1em}
    \item \textit{Socio-economic status.} Education, occupation, and social class are classic dimensions on which people sort in social networks \citep{mcpherson2001birds}. Since our subjects are all students, we include two measures of family background. First, as in \cite{conti2013popularity}, we consider \textit{parental education}, measured for individual $i$ as the average education between the mother and the father and included for a pair $i$ and $j$ in the analysis as the absolute difference in the average parental education between $i$ and $j$. Second, we consider the absolute difference between the self-reported \textit{household income} of $i$ and $j$.

\item \textit{Common interests.} Students are more likely to become friends if they share interests, as documented in \citep{marmaros2002peer}. Therefore, we include two variables related to preferences regarding the field of study. First, \textit{STEM best} is a dummy equal to 1 if  $i$ and $j$ have received the highest grade in their final high-school exam in the scientific track (that includes topics from science, technology, engineering, or mathematics (STEM) subjects), rather than in humanities or social sciences. Second, \textit{STEM preference} is a dummy variable equal to 1 if both individuals report a STEM subject as the most preferred, irrespective of their grades.
    
\item \textit{Prosocial attitudes.} There exists large evidence that prosocial attitudes correlate with network positioning, including with who befriends whom \citep{girard2015, branas2010altruism, apicella2012social, kovavrik2012}. We first use the behavior in the Dictator game (DG) elicited together with the networks in Periods 1 and 2 to measure subjects' \textit{altruism}. DG is a classic experimental protocol, in which a ``Dictator" splits a fixed amount of money between herself and another anonymous individual, the Recipient. Dictators rarely keep all the money for themselves; on average, 20\% of the money is donated to the recipient, despite the anonymity of the transfers. The donated quantity is usually interpreted as an instance of altruistic behavior. As a measure of altruism, we employ the amount donated in the DG in Period 2, which--we believe--reflects better the social dynamics in the groups under study. Once again, the variable included in the regressions is the absolute difference between the altruism exhibited by $i$ and $j$. In addition, since the amount donated in the DG by an individual generally differs across the two periods and, in general, decreased between the two time periods, we define \textit{altruism learning} as equal to one if the amount given in Period 2 is lower than in Period 1 for both $i$ and $j$ (and zero otherwise). This variable enables to test whether there can be homophily in learning. We additionally measure \textit{reciprocity} from the answer to the following survey question: ``I am willing to do a boring job to return previous help''. The answers vary from 1 (minimum willingness to help) to 7 (maximum reciprocity) and we construct the variable \textit{reciprocity} between $i$ and $j$ by taking the absolute difference between their answers to the survey question. Last, \textit{volunteering} was elicited from the answer to the question ``Are you a member of a voluntary organization (e.g., Red Cross, NGO, Political Party, Sports Club, Church Choir, economic association,...)?''. We assess homophily in volunteering using a dummy equal to 1 if both $i$ and $j$ answered yes to this question. 

\item \textit{Self-confidence.} One's self-confidence is an important determinant of her socio-economic outcomes in a range of domains \citep{sharot2011optimism}. We measure \textit{self-confidence} from the answer to the question ``When I face a problem I usually trust that I will solve it'', the answer could take values from 1 (minimum self-confidence) to 7 (maximum self-confidence). Homophily in self-confidence is captured by the absolute difference between both individuals’ self-confidence scores.
    
\item \textit{Risk and time preferences.} Common attitudes to risk and delays may affect who befriends whom. Indeed, \cite{kovarik2014risk} document a correlation between the clustering coefficient and risk aversion in college friendship networks and \cite{girard2015} detects homophily on time but not risk preferences. In our data, both \textit{risk preferences} and \textit{time preferences} were elicited using multiple price lists (see  \cite{andersen2006elicitation} for a review). As for risk, participants made 10 binary decisions between two lotteries with different amounts and probabilities. This elicitation was incentivized. In case of time preferences, they faced 11 binary decisions between a lower immediate or earlier amount of money and a later but higher quantity.\footnote{The elicitation of time preferences was hypothetical. However a recent paper shows that hypothetical and incentivized time preferences are the same (see \cite{BranasTIME}).}  Following the literature, we use the number of times an individual selects the less risky option and the number of decisions, in which she chooses the immediate or earlier amount, as the measures of risk aversion and impatience, respectively. In both cases, homophily in preferences is reflected in the absolute difference between both individuals' risk and time scores.

\item \textit{Cognitive characteristics.} Students in the same class frequently engage in group activities and sorting on intelligence is probably the earliest evidence of homophily in the literature \cite{almack1922influence}. More recently, \cite{conti2013popularity} suggest that IQ and other cognitive variables might condition the formation of social networks or groups in certain contexts, \cite{kretschmer2018selection} show that students tend to befriend peers with similar academic achievement. We employ two measures of cognitive abilities in our analysis. First, we generate a variable \textit{reflective} taking value of 1 if two individuals scored a positive amount in the Cognitive Reflection Test by \cite{frederick2005cognitive} -- see \cite{BranasCRT} for a meta-analysis. Second, we include the absolute difference in CRT between $i$ and $j$, labeled as \textit{CRT}. Both variables reflect whether two individuals have a similar cognitive style, but \textit{CRT} captures it more precisely. 

Finally, \cite{chapman2018loss} estimate that choice consistency in tasks similar to our risk and time preference elicitation procedure is positively correlated with cognitive abilities, education, and income. Therefore, we generate a variable \textit{inconsistency} and set it equal to 2 if an individual exhibits inconsistent preferences in both risk and time preferences (i.e., switches between the risk averse/patient to risk loving/impatient options more than once; see \cite{andersen2006elicitation}  or \cite{BranasEER} for a recent review), equal to 1 if she exhibit inconsistent preferences either in the risk or time elicitation, and 0 if she answer the task consistently. Again, we compute homophily in inconsistency as the absolute difference between both individuals’ inconsistency scores.
    
\item \textit{Political attitudes.} Common political orientation has shown to be correlated with friendship formation  \cite{mcpherson2001birds, marmaros2006friendships, mayer2008old}. We elicit political orientation of the individual from the answer to the question ``Do you think the size of the public sector in Spain is too large?'' and define a dummy variable \textit{political orientation} equal to one if both individuals state a positive answer. It is a indirect measure of political orientation. 
\end{enumerate}    
\end{itemize}

\begin{table}[h!]
\footnotesize
\centering
\caption{Potential determinants of Friendship}
\begin{tabular}{lll} \hline
Variable & Transformation & Description\\\hline
Observable  & & \\\hline
Gender  &  D & 1 = Both individuals are of the same gender.\\
Smoking & D & 1 = Both individuals are smokers. \\
Common section & D & 1 = Both individuals in the same section.\\
BMI & A & Absolute difference between the BMI of $i$ and $j$.\\
Friends Period 1 & D & 1= If $i$ and $j$ were friends in Period 1.\\
\hline
Unobservable  & & \\\hline
Parental education & A & Absolute difference between the average parental education of $i$ and $j$.\\
STEM best & D & 1 = Both individuals obtained their highest mark in a STEM subject.\\
STEM preference & D & 1 = Both individuals selected a STEM subject as their most preferred \\
& & subject.\\
Reflective & D & 1 = Both individuals scored more than 0 in the CRT. \\
CRT & A & Absolute difference in cognitive reflection test scores between $i$ and $j$. \\
Household Income & A & Absolute difference between the household income of $i$ and $j$.\\
Altruism learning & D & 1= Both individuals give less in the dictator game of period 2 than\\ & &  in period 1.\\
Altruism & A & Absolute difference between individuals’ amount given in the dictator \\  & &  game of period 2.\\
Reciprocity & A & Absolute difference between individuals’ willingness to return help, \\ & & previously given.\\
Kindness & D & 1= Both individuals were members of a voluntary organization.\\
Self-confidence & A & Absolute difference between both individuals’ self-confidence score.\\
Risk preferences & A & Absolute difference between both individuals’ risk score.\\
Time preferences & A & Absolute difference between both individuals’ patience score.\\
Inconsistency & A & Absolute difference between both individuals’ inconsistency score.\\
Political orientation & D & 1= Both individuals have a `right-wing' political orientation.\\
\hline

\end{tabular}
\begin{minipage}{16cm}
   \footnotesize Transformation=D if a variable is a dummy equal to 1 if $i$ and $j$ share the same characteristic; Transformation=A if the variable is ordinal and the absolute difference between the characteristics of $i$ and $j$ employed.
  \end{minipage}
\label{Table varlist}
\end{table}

Table \ref{Table ss} summarizes all the individual characteristics in our sample. We observe that 44\% of the participants are female, 22\% are smokers, 44\% belief that the public sector is too large (suggesting relatively more right-wing political orientation), and 30\% state a science-oriented (STEM) fields as the most preferred. We also note that the average CRT and altruism are relatively low, 0.55 and 1.07, respectively. On the other hand, the average self-confidence index is high, 5.7. 

\begin{table}[h!]
\centering
\caption{Summary statistics}
\begin{tabular}{lccccc} \hline
Variable & N & Mean & Std. Dev. & Min. & Max.\\\hline
Observable & & & \\\hline
Female & 244 &  0.44 & 0.50 & 0 & 1\\
Smoking & 186 & 0.22 & 0.42 & 0 & 1\\
BMI & 186& 22.53 & 3.23 & 15.60& 41.87\\
Friends Period 1 & 204 & 2.7& 2.1& 0& 11\\
Friends Period 2 & 246 & 7.3&4.5 & 0& 18\\
\hline
Unobservable & & & \\\hline
Parental education & 186 & 2.73 & 1.53 & 0 & 5\\
STEM best & 217 & 0.28 & 0.45 & 0 & 1\\
STEM preference & 217 & 0.30 & 0.46 & 0 & 1\\
Reflective & 274 & 0.54 & 0.50 & 0 & 1\\
CRT & 217 & 0.55 & 0.73 & 0 & 3 \\
Household Income & 185 & 2.39 & 1.21 & 0 & 5\\
Altruism learning & 274 & 0.38 & 0.49 & 0 & 1\\
Altruism & 168 & 1.07 & 1.06 & 0 & 5\\
Reciprocity & 186 & 3.57 & 1.81 & 1 & 7\\
Volunteering & 186 & 0.17 & 0.38 & 0 & 1\\
Self-confidence & 186 & 5.74 & 1.23 & 1 & 7\\
Risk preferences & 195 & 4.94 & 1.43 & 0 & 10\\
Time preferences & 168 & 5.60 & 2.54 & 0 & 11\\
Inconsistency & 202 & 0.15 & 0.36 & 0 & 1\\
Political orientation (right)& 186 & 0.44 & 0.50 & 0 & 1\\
\hline
\end{tabular}
\begin{minipage}{12cm}
  \end{minipage}
\label{Table ss}
\end{table}

\subsection{Friendship model}
\label{sec: model}
Our aim is to detect the main determinants of frienship formation and estimate their importance in the friendship formation process. The specification of our friendship model relies on homophily theories, positing that individuals prefer to befriend with individuals who are similar to them. As a result, we expect that two individuals are more likely to be friends if they share personal characteristics. However, there is no agreement in the literature about the importance of the different characteristics in explaining link formation nor whether observable characteristics, such as gender or race, should matter more than features invisible to the naked eye, such as cognitive or personality traits. As a result, \textit{a priori}, we have no specific empirical model and little guidance as to which factors are likely to be influential in the friendship formation process. We deal with this problem, known as \textit{model uncertainty} in the literature, using a Bayesian Model Averaging (BMA) approach \cite{hoeting1999bayesian}. In this section, we present the friendship formation model and the BMA methodology.

Our friendship model starts with a basic pairwise probability model that relates the probability that individuals $i$ and $j$ are friends in Period 2, may 2011, with their similarity in the characteristics presented and explained in Section \ref{sec: var}, $X_{ij}$, conditioning on the probability of being friends in Period 1. Such specification follows standard approaches of modelling network formation in the literature (see e.g. the friendship formation in \cite{marmaros2006friendships, mayer2008old} or the co-authorship model of \cite{fafchamps2010matching}). Formally,
\begin{equation}
    P(g_{ij,2}=1/X_{ij})=\beta_{0}+\beta_{1}|X_{i}-X_{j}|+\beta_{2} D_{ij}+\beta_{3} g_{ij,1}+u_{ij},
\label{eq: friend}    
\end{equation}
where $|X_{i}-X_{j}|$ is the vector of absolute differences between the non-binary measures of individuals $i$ and $j$, while $D_{ij}$ is a vector containing all dummy variables that capture common dichotomous characteristics as specified in Table \ref{Table varlist}. The variable $g_{ij,t}=1(0)$ if $i$ and $j$ were (not) friends in Period $t \in \{1,2\}$ and $u_{ij}$ is the disturbance term. Below, we also estimate this model for females and males separately to analyse whether the friendship determinants differ across genders.

\subsection{Methodology}
\label{sec: bma}
We consider two different approaches to identify the underlying factors explaining link formation in the network: BMA and, mostly for robustness, Weighted average least squares (WALS; see Section \ref{sec: rob}). 

Our benchmark estimation method is the BMA reported in Section \ref{sec: findings}. BMA is specifically designed to address model uncertainty by estimating model (\ref{eq: friend}) for all possible combinations of the regressors and taking a weighted average over all the candidate models, where the weights are determined by applying Bayes' rule. In simple terms, BMA only selects the determinants of link formation that are not too sensitive to the inclusion or removal of other factors into the model specification. Therefore, it explicitly accounts for the omitted-variable and multicollinearity issues discussed in Section \ref{sec: intro}. The probability that model $j$, $M_j$, is the "true" model given the data, $y$ (i.e., the posterior model distribution given a prior model probability) is defined as
\begin{equation}
P(M_j|y)=\frac{P(y|M_j)P(M_j)}{\Sigma_{i=1}^{2^k}P(y|M_i)P(M_i)},
\end{equation}
where $P(y|M_j)$  is the marginal likelihood of Model $j$, $P(M_i)$ is the prior model probability, and $\Sigma_{i=1}^{2k}P(y|M_i)P(M_i)$ is the integrated likelihood of model $j$. We employ the priors specified in \cite{magnus2010comparison}. In particular, they consider uniform priors on the model space, so each model has the same probability of being the true one. Moreover, they use a Zellner's g-prior structure for the regression coefficients and set the hyperparameter $g=\frac{1}{max(N,K^{2})}$ as in Fernandez et al. (2001), where $K$ is the number of regressors and $N$ the number of observations. This hyperparameter measures the degree of prior uncertainty over the coefficients. For robustness, we also consider 'random theta', 'fixed' and 'uniform' priors for the model space. 

In the next section, we present the estimates of the \textit{posterior inclusion probability (PIP)} of an explanatory factor, which can be interpreted as the probability that a particular regressor belongs to the true model of friendship. We also present results for the posterior mean, the coefficients averaged over all models, and the posterior standard deviation, which describes the uncertainty in the parameters and the model. In the robustness section, we will present results of the WALS (Section \ref{sec: rob}), a recent model average approach, which takes an intermediate position between frequentist and Bayesian methods.

\section{Results}
\label{sec: findings}
Table $\ref{Table BMAall}$ reports our main estimation results, obtained using the BMA approach described in Section \ref{sec: bma} over the 168 students who participated in all the surveys and economic games troughout the whole academic year. The table lists all the determinants of friendship formation under scrutiny ranked by their estimated inclusion probability ($PIP$). The first column lists the different dimensions of homophily and the second column presents the $PIP$ of each potential determinant of friendship formation in May 2011. As a rule of thumb, a factor is considered a very robust predictor of friendship if $PIP \geq 0.80$.\footnote{This is a relatively conservative threshold as compared to the literature, but note from Tables \ref{Table BMAall} - \ref{Table BMAmale} that all our conclusions hold for any threshold $PIP \geq 0.3$.} 

We find that the most robust determinants for the full sample are the variables capturing common gender, common smoking habits, belonging to the same section, and being already friends before entering into the college. The only observable dimension under study that does not predict friendship robustly is BMI.

Moreover, all the robust determinants affect friendship with the expected sign (see the posterior mean in the third column of Table $\ref{Table BMAall}$). More precisely, people are 1.7 percentage points more likely to befriend each other if they have the same gender. Smoking is even a stronger driver of friendship than gender: if two individuals smoke, they have a 3.4 percentage points higher probability of forming a tie in our network. This is consistent with \cite{christakis2008collective}, \cite{hsieh2018} and \cite{lorant2019peer} who report evidence of strong peer effects in the uptake of smoking. Last, although people were assigned into the study sections randomly, being assigned to the same section plays a important role in link formation of students' networks, increasing the probability of being friends by 6.4\%. This corroborates the existing evidence regarding the key role of meeting opportunities in network formation. Unsurprisingly, being friend in October 2010 is the strongest predictor of friendship in May 2011. 

\begin{table}[htp!]
\caption{Determinants of link formation in students' network}
\centering
\vspace{0.5cm}
\begin{tabular}[c]{lccc}\hline
\\
& PI prob. & Pt. Mean & Pt. Std. \\ \hline
\\
\textbf{Common Gender}&1& 0.017 & 0.003\\[4pt]
\textbf{Common Section}&1& 0.064 &0.003\\[4pt]
\textbf{Friends$_{t-1}$} &1&0.435&0.023\\[4pt]
\textbf{Both Smokers}&0.97&0.034&0.010\\[4pt]
Inconsistent diff. & 0.22 & -0.022& 0.005\\[4pt]
Altruism diff.&0.08&-0.003&0.001\\[4pt]
CRT diff. & 0.04 & -0.0001 & 0.0008\\[4pt]
Both Reflective &0.03&0.0002&0.011\\[4pt]
Time pref. diff. &0.02&-0.0000&0.0001\\[4pt]
Income diff. &0.02&-0.0000&0.0003\\[4pt]
Risk diff.&0.01& 0.0000 &0.0001\\[4pt]
Reciprocity diff.&0.01& 0.0000 &0.0002\\[4pt]
Self-confidence diff.&0.01& -0.0000 &0.0002\\[4pt]
BMI diff.&0.01& -0.0000 &0.0001\\[4pt]
Parent educ. diff.&0.01& -0.0000 &0.0001\\[4pt]
Both volunteers&0.01& -0.0000 &0.0012\\[4pt]
Both STEM best grade&0.01& 0.0000 &0.0006\\[4pt]
Both STEM pref. &0.01& 0.0000 &0.0006\\[4pt]
Both altruism learner&0.01& -0.0000 &0.0003\\[4pt]
Both right&0.01& 0.0000 &0.0004\\[4pt]
Observations & 8515 & 8515& 8515\\
\hline
\\
\end{tabular}
\begin{minipage}{14cm}
   \footnotesize  \textit{Notes}: Column 1 presents the posterior inclusion probability ($PIP$). Robust determinants are those with $PIP \geq 0.8$ in bold. Column 2 shows the posterior mean. Column 3 reports the posterior standard deviation. The dependent variable is equal to 1 if $i$ and $j$ are friends in Period 2, 0 otherwise. The results are obtained by using a uniform prior for the prior model probability and a BRIC prior for the hyperparameter that measures the degree of prior uncertainty on coefficients, $g=1/max(N,K^{2})$.
  \end{minipage}
\label{Table BMAall}
\end{table}		

In contrast to the observable dimensions, none of the imperceptible characteristics matter for friendship formation in our data. Inconsistent risk and time preferences are the best predictor with $PIP=0.22$, relatively low inclusion probability that cannot rival with the $PIP$'s estimated for the observable features. The remaining unobservable characteristics exhibit an estimated $PIP < 0.1$.

Hence, the first conclusions of our analysis is that, if we focus on the whole sample under study, only observable dimensions matter for friendships, while none of the unobservable factors play an robust role in the friendship formation process in our data.

To understand whether the friendship formation processes differ across genders, we reestimate the friendship model separately for the male and female participants. The results are presented in Tables \ref{Table BMAfemale} and \ref{Table BMAmale} in the appendix, respectively. In line with the pooled estimates, common gender, smoking habits, and belonging to a common section are important drivers of friendship for women; these factors have the associated $PIP=1$. 

In a stark contrast, the only robust predictors determining friendship formation in men is belonging to the same study section that increases the probability of forming a link by 6.8\% with $PIP=1$.

The gender differences notwithstanding, the gender-specific estimates corroborate that observable characteristics are important drivers of friendship formation while imperceptible features play a minor role. 

\section{Robustness}
\label{sec: rob}
In this section, we check the robustness of the results to (i) the employment of different priors in the BMA model, (ii) applying a different estimation method designed to identify the most robust drivers of friendship formation, and (iii) removing the students who have already had friends in Period 1 (October 2010). The first two tests check whether our results are not a consequence of our estimation procedure, while the third case allows to see whether our results are not driven by preexisting social ties in the group. 

\begin{figure}[htp!]
\caption{Determinants of link formation in students' network: PIP using different model priors}
\label{fig: BMA}
\begin{centering}
\begin{tabular}{cc}
\includegraphics[scale=0.30, angle=0]{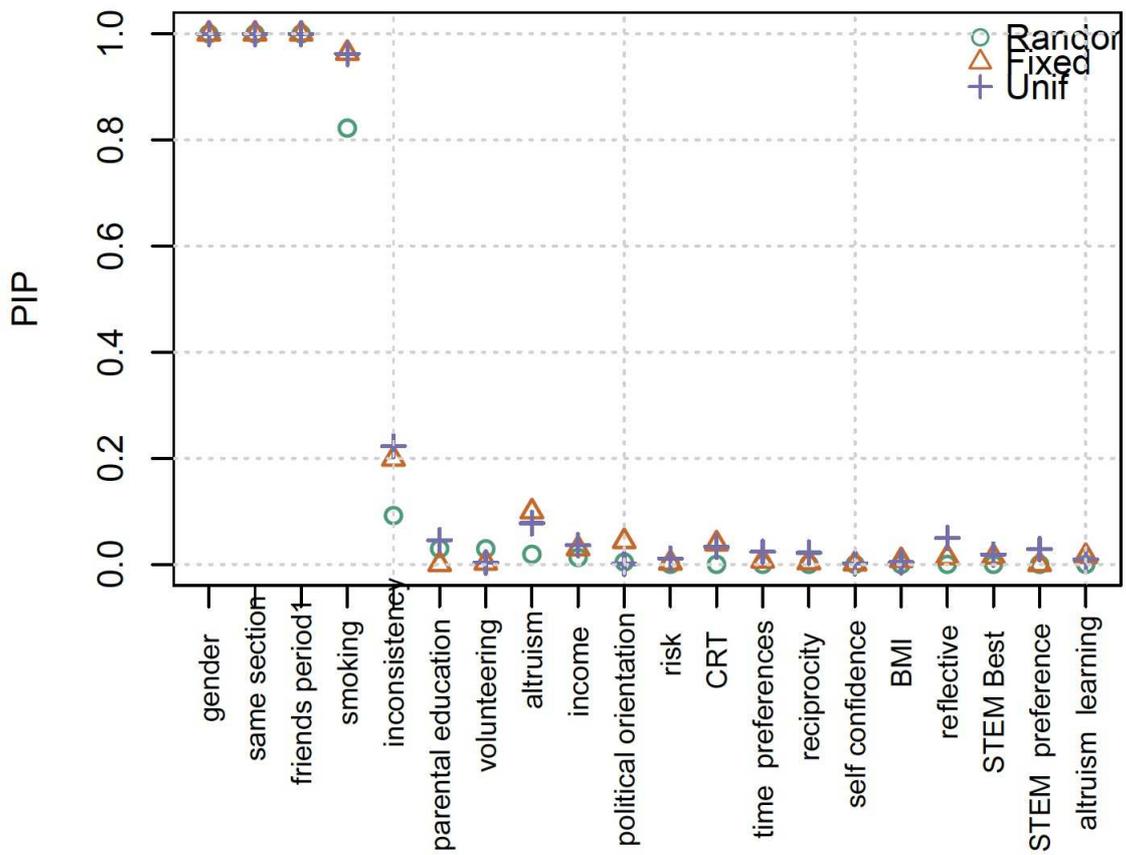}  \\
\end{tabular}
\par\end{centering}
\noindent {\footnotesize{Note: The random prior corresponds to the 'random theta' prior by \cite{ley2009effect}, who suggest a binomial-beta hyperprior on the a priori inclusion probability. The fixed is the fixed common prior inclusion probabilities for each regressor as in \cite{sala2004determinants}. Unif corresponds to the uniform model prior. The sample includes 8,515 observations.}}

\end{figure}

As for (i), we present results for an analysis using two alternative priors for the model probability: the ``random theta'' prior proposed by \cite{ley2009effect} who advocate for a \textit{binomial-beta hyperprior} on the \textit{a priori inclusion probability}; and \textit{fixed common prior inclusion probabilities} for each regressor as in \cite{sala2004determinants}. Figure \ref{fig: BMA} shows that our main findings are largely robust to the specification of the model priors: the most important drivers of link formation in our data are the same regardless of the prior specification. 

Second, we test whether the findings hold while using alternative methods dealing with model uncertainty. To that end, we apply the WALS method introduced by \cite{magnus2010comparison}. The rule of thumb with this method is that an explanatory factor is considered robust if the absolute value of the $t$-statistic is higher than 2. The results for the full sample, presented in Table \ref{Table WALS}, show that the most robust drivers are the same as in Table \ref{Table BMAall}. Nonetheless, this alternative method suggest an additional determinant of friendship formation: \textit{altruism} with a $t$-statistic of -2.79. The weighted average coefficient of the absolute difference of the amount donated in the DG between both individuals is 0.004, implying that homophily on altruism is associated with an increase of the probability that two individuals befriend each other by 0.4 percentage points. This does not affect our conclusions that observable features are considerably more relevant than imperceptible characteristics, but it suggests that some of the latter should still be considered and future work should determine which observable features robustly predict tie formation across contexts and cultures (see the next section for a discussion of this and other issues).

\begin{table}[h!]
\centering
\caption{Determinants of link formation in students' network: A Weighted Average Least Squares approach (WALS). Full sample.}
\vspace{0.5cm}
\begin{tabular}[c]{lccc}\hline
\\
& Coef. & Std. & t-stat. \\ \hline
\\
\textbf{Common Section}&0.059& 0.003 & 20.20\\[4pt]
\textbf{Friends$_{t-1}$} & 0.420&0.021& 19.81\\[4pt]
\textbf{Common Gender}& 0.016& 0.003 & 6.04\\[4pt]
\textbf{Both Smokers}&0.031&0.008& 3.89\\[4pt]
\textbf{Altruism diff.} &-0.004& 0.001& -2.79\\[4pt]
CRT diff. & -0.003 & 0.002 & -1.81\\[4pt]
Inconsistent diff. & -0.007 & 0.004 & -1.75\\[4pt]
Both Reflective & 0.005&0.003 & 1.47\\[4pt]
Parent educ. diff.&-0.001& 0.0010 & -1.18\\[4pt]
Reciprocity diff.&0.0008& 0.0008 & 1.01\\[4pt]
Time pref. diff. &-0.0005& 0.0006 & -0.81\\[4pt]
Both altruism learner& -0.0018& 0.0026 & -0.71\\[4pt]
Both right&0.002& 0.0033 &0.59\\[4pt]
Risk diff.&0.0005& 0.0009 &0.57\\[4pt]
Income diff. & -0.0006& 0.0013&-0.43\\[4pt]
Both volunteers&-0.0034& 0.0087&0.39\\[4pt]
BMI diff.&0.0002& 0.0005 &0.32\\[4pt]
Both STEM pref. &0.0007& 0.0044 &0.16\\[4pt]
Self-confidence diff.&-0.0001& 0.0012 & -0.05\\[4pt]
Both STEM best grade&-0.0001& 0.0051 &-0.03\\[4pt]
Observations & 8515 & 8515& 8515\\
\hline
\\
\end{tabular}
\begin{minipage}{12.1cm}
   \footnotesize  \textit{Notes}: The results are obtained by using the Weighted Average Least Squares approach introduced by \cite{magnus2010comparison}. Determinants with a $t > 2$ are considered robust. 
\end{minipage}
\label{Table WALS}
\end{table}	

Finally, to focus on the pure formation process (free of any pre-college influences), we reestimate our benchmark model excluding all the students who have already had at least one friend in the first week of their first academic year (Period 1). This decreases the number of potential pairs to 8483 observations, but the results presented in Table \ref{Table BMAall_es} are both quantitatively and qualitatively similar to those obtained using the full sample.

\begin{table}[h!]
\caption{Determinants of link formation in students' network: Excluding students with friends in period 1.}
\centering
\vspace{0.5cm}
\begin{tabular}[c]{lccc}\hline
\\
& PI prob. & Pt. Mean & Pt. Std. \\ \hline
\\
\textbf{Common Gender}&1& 0.015 & 0.003\\[4pt]
\textbf{Common Section}&1& 0.064 &0.003\\[4pt]
\textbf{Both Smokers}&0.87&0.026&0.013\\[4pt]
Inconsistent diff. & 0.07 & -0.0006& 0.0023\\[4pt]
Altruism diff.&0.06&-0.0002&0.0007\\[4pt]
CRT diff. & 0.05 & -0.0002 & 0.0008\\[4pt]
Time pref. diff. &0.03&-0.0000&0.0002\\[4pt]
Both Reflective &0.03&0.0001&0.0010\\[4pt]
Both volunteers&0.02& -0.0001 &0.0016\\[4pt]
Income diff. &0.02&-0.0000&0.0002\\[4pt]
Risk diff.&0.01& 0.0000 &0.0001\\[4pt]
Reciprocity diff.&0.01& 0.0000 &0.0001\\[4pt]
Self-confidence diff.&0.01& -0.0000 &0.0002\\[4pt]
BMI diff.&0.01& 0.0000 &0.0001\\[4pt]
Parent educ. diff.&0.01& -0.0000 &0.0001\\[4pt]
Both STEM best grade&0.01& 0.0000 &0.0006\\[4pt]
Both STEM pref. &0.01& 0.0000 &0.0006\\[4pt]
Both altruism learner&0.01& -0.0000 &0.0003\\[4pt]
Both right&0.01& 0.0000 &0.0005\\[4pt]
Observations & 8483 & 8483& 8483\\
\hline
\\
\end{tabular}
\begin{minipage}{14cm}
   \footnotesize  \textit{Notes}: Column 1 presents the posterior inclusion probability ($PIP$). Robust determinants are those with $PIP \geq 0.8$ in bold. Column 2 shows the posterior mean. Column 3 reports the posterior standard deviation. The dependent variable is equal to 1 if $i$ and $j$ are friends in Period 2, 0 otherwise. The results are obtained by using a uniform prior for the prior model probability and a BRIC prior for the hyperparameter that measures the degree of prior uncertainty on coefficients, $g=1/max(N,K^{2})$.
  \end{minipage}
\label{Table BMAall_es}
\end{table}		


\section{Discussion}
\label{sec: discussion}

In this study, we argue that the literature has proposed a large array of dimensions of homophily without accounting for the model uncertainty problem inherent in the identification of the determinants of friendship formation. This research aims at initializing such a research line. Focusing on 20 particular individual characteristics that cover a large share of the determinants of segmentation of heterogeneous populations analyzed so far, we find that the robust determinants of friendship formation are gender, smoking habits, and geographical/spatial proximity. Hence, our data suggest that, in our context of an emerging social structure, all robust predictors of friendships are directly visible by a naked eye. Due to the homogeneity of our sample, we do not consider other observable attributes such as race, ethnicity, and religion, but the literature shows that these characteristics are key predictors of links in many contexts \citep{lazarsfeld1954friendship, ibarra1995race, mcpherson2001birds, moody2001race, marmaros2006friendships, currarini2010identifying}. We thus positions them among the robust predictors of friendships generally. In contrast, we find little evidence for unobservable features to predict who befriends whom in an emerging social network among University freshmen. Neither political orientation, family background, nor cognitive and behavioral traits survive our robustness tests.


Our findings corroborate some existing findings and contradict others. First, gender belongs among the most robust findings and we are no exception. Since gender segregation might have contributed to e.g. the gender wage and output gaps \citep{ductor2018gender, lindenlaub2019}, we corroborate that policies aiming at a large gender integration in social and professional networks could reduce gender inequalities. As for smoking, we corroborate \cite{hsieh2018} and \cite{christakis2008collective} that smoking influences friendship formation. However, our results suggest that this factor is more important for women. If corroborated in other studies, this finding calls for gender-specific policies combating smoking. Last, we confirm that friendship formation can be to a large extend malleable and shaped by policy interventions as random distribution of people into groups have large consequences on who befriends whom (see also \cite{marmaros2006friendships, girard2015}).


In contrast, the fact that the unobservable factors under scrutiny, like personality, economic preferences, family background, political orientation are not robust predictors of link formation in the students' networks contradict others studies (e.g. \cite{almack1922influence, marmaros2006friendships, christakis2007spread, mayer2008old, girard2015}). We do not argue that none of them is correlated with friendship. Rather, we argue that whether they can predict friendship depends on other covariates in the link formation model, suggesting that the unobservable, imperceptible attributes are not the primary reason for how people become friends.

Our results have implications for several strands of network literature. First, it is well documented that friendship networks have far-reaching and long-lasting socio-economic consequences in human life (see \cite{burt1997contingent, christakis2007spread, christakis2008collective, cohen2008small, conti2013popularity, jackson2017economic} for a few examples of the large literature on the topic). We contribute to this literature by emphasizing which individual attributes robustly predict who befriends whom. Understanding the determinants of friendship and segregation patterns is crucial to evaluate diffusion of information, behaviors, and diseases in general populations and to design optimal policies aimed at maximizing or minimizing the diffusion, targeting integration of minorities, and combating gender and race inequalities.

Second, our results deliver an important message for the estimation of network effects. One of the main empirical challenges for causal estimation of network effects is to account for all factors that influence friendship and network formation \citep{manski1993identification, boucher2016some, jackson2017economic}. Nevertheless, many individual characteristics are typically not observed by the researchers or policy-makers and they are believed to influence both the outcomes under study and the link formation. These concerns have motivated a series of methodological contributions enabling the estimation of network effects causally, each under different assumptions (examples include \cite{bramoulle2009identification, de2010identification, hsieh2016social}). It is commonly assumed that people exhibit homophily in the unobserved attributes. Our results suggest that such concerns might be less serious than expected and that one can account for the endogeneity of the network structure by controlling for observable personal attributes, at least in the context of friendship networks. Future research should determine to what extent our results replicate in other data and contexts.  

Third, we may also indirectly inform the literature on the origins of homophily by testing the relevance of different classes of attributes in friendship formation. In particular, our results suggest that observable characteristics might be the primary driving force behind homophilous interactions, while unobservable, imperceptible features might be solely a by-product of the many correlations between the observable and imperceptible attributes. Naturally, this claim requires further analysis using different data, contexts, and definitions of links before one can widely accept its validity.

\clearpage
\phantomsection\addcontentsline{toc}{section}{Bibliography}{
\bibliographystyle{chicago}
{\footnotesize \setstretch{0} \bibliography{Homophily}}

\begin{thebibliography}{}

\bibitem[\protect\citeauthoryear{Almack}{Almack}{1922}]{almack1922influence}
Almack, J.~C. (1922).
\newblock The influence of intelligence on the selection of associates.
\newblock {\em School and Society\/}~{\em 16\/}(410), 529--530.

\bibitem[\protect\citeauthoryear{Altonji and Blank}{Altonji and
  Blank}{1999}]{altonji1999race}
Altonji, J.~G. and R.~M. Blank (1999).
\newblock Race and gender in the labor market.
\newblock {\em Handbook of Labor Economics\/}~{\em 3}, 3143--3259.

\bibitem[\protect\citeauthoryear{Amador-Hidalgo, Brañas-Garza, Espín,
  García-Muñoz, and Hernández-Román}{Amador-Hidalgo
  et~al.}{2021}]{BranasEER}
Amador-Hidalgo, L., P.~Brañas-Garza, A.~M. Espín, T.~García-Muñoz, and
  A.~Hernández-Román (2021).
\newblock Cognitive abilities and risk-taking: Errors, not preferences.
\newblock {\em European Economic Review\/}~{\em 134}, 103694.

\bibitem[\protect\citeauthoryear{Andersen, Harrison, Lau, and
  Rutstr{\"o}m}{Andersen et~al.}{2006}]{andersen2006elicitation}
Andersen, S., G.~W. Harrison, M.~I. Lau, and E.~E. Rutstr{\"o}m (2006).
\newblock Elicitation using multiple price list formats.
\newblock {\em Experimental Economics\/}~{\em 9\/}(4), 383--405.

\bibitem[\protect\citeauthoryear{Apicella, Marlowe, Fowler, and
  Christakis}{Apicella et~al.}{2012}]{apicella2012social}
Apicella, C.~L., F.~W. Marlowe, J.~H. Fowler, and N.~A. Christakis (2012).
\newblock Social networks and cooperation in hunter-gatherers.
\newblock {\em Nature\/}~{\em 481\/}(7382), 497--501.

\bibitem[\protect\citeauthoryear{Barber and Odean}{Barber and
  Odean}{2001}]{barber2001boys}
Barber, B.~M. and T.~Odean (2001).
\newblock Boys will be boys: Gender, overconfidence, and common stock
  investment.
\newblock {\em The Quarterly Journal of Economics\/}~{\em 116\/}(1), 261--292.

\bibitem[\protect\citeauthoryear{Boucher and Fortin}{Boucher and
  Fortin}{2016}]{boucher2016some}
Boucher, V. and B.~Fortin (2016).
\newblock Some challenges in the empirics of the effects of networks.
\newblock {\em Handbook on the Economics of Networks\/}.

\bibitem[\protect\citeauthoryear{Bramoull{\'e}, Currarini, Jackson, Pin, and
  Rogers}{Bramoull{\'e} et~al.}{2012}]{bramoulle2012homophily}
Bramoull{\'e}, Y., S.~Currarini, M.~O. Jackson, P.~Pin, and B.~W. Rogers
  (2012).
\newblock Homophily and long-run integration in social networks.
\newblock {\em Journal of Economic Theory\/}~{\em 147\/}(5), 1754--1786.

\bibitem[\protect\citeauthoryear{Bramoull{\'e}, Djebbari, and
  Fortin}{Bramoull{\'e} et~al.}{2009}]{bramoulle2009identification}
Bramoull{\'e}, Y., H.~Djebbari, and B.~Fortin (2009).
\newblock Identification of peer effects through social networks.
\newblock {\em Journal of Econometrics\/}~{\em 150\/}(1), 41--55.

\bibitem[\protect\citeauthoryear{Branas-Garza, Cobo-Reyes, Espinosa,
  Jim{\'e}nez, Kov{\'a}{\v{r}}{\'\i}k, and Ponti}{Branas-Garza
  et~al.}{2010}]{branas2010altruism}
Branas-Garza, P., R.~Cobo-Reyes, M.~P. Espinosa, N.~Jim{\'e}nez,
  J.~Kov{\'a}{\v{r}}{\'\i}k, and G.~Ponti (2010).
\newblock Altruism and social integration.
\newblock {\em Games and Economic Behavior\/}~{\em 69\/}(2), 249--257.

\bibitem[\protect\citeauthoryear{Branas-Garza, Kov{\'a}{\v{r}}{\'\i}k, and
  Neyse}{Branas-Garza et~al.}{2013}]{branas2013second}
Branas-Garza, P., J.~Kov{\'a}{\v{r}}{\'\i}k, and L.~Neyse (2013).
\newblock Second-to-fourth digit ratio has a non-monotonic impact on altruism.
\newblock {\em PLoS ONE\/}~{\em 8\/}(4), e60419.

\bibitem[\protect\citeauthoryear{Brañas-Garza, Jorrat, Espín, and
  Sánchez}{Brañas-Garza et~al.}{2020}]{BranasTIME}
Brañas-Garza, P., D.~Jorrat, A.~M. Espín, and A.~Sánchez (2020).
\newblock Paid and hypothetical time preferences are the same: Lab, field and
  online evidence.
\newblock Technical report, arXiv:2010.09262.

\bibitem[\protect\citeauthoryear{Brañas-Garza, Kujal, and
  Lenkei}{Brañas-Garza et~al.}{2019}]{BranasCRT}
Brañas-Garza, P., P.~Kujal, and B.~Lenkei (2019).
\newblock Cognitive reflection test: Whom, how, when.
\newblock {\em Journal of Behavioral and Experimental Economics\/}~{\em 82},
  101455.

\bibitem[\protect\citeauthoryear{Burt}{Burt}{1997}]{burt1997contingent}
Burt, R.~S. (1997).
\newblock The contingent value of social capital.
\newblock {\em Administrative Science Quarterly\/}, 339--365.

\bibitem[\protect\citeauthoryear{Carrell, Sacerdote, and West}{Carrell
  et~al.}{2013}]{carrell2013natural}
Carrell, S.~E., B.~I. Sacerdote, and J.~E. West (2013).
\newblock From natural variation to optimal policy? the importance of
  endogenous peer group formation.
\newblock {\em Econometrica\/}~{\em 81\/}(3), 855--882.

\bibitem[\protect\citeauthoryear{Centola}{Centola}{2011}]{centola2011experimental}
Centola, D. (2011).
\newblock An experimental study of homophily in the adoption of health
  behavior.
\newblock {\em Science\/}~{\em 334\/}(6060), 1269--1272.

\bibitem[\protect\citeauthoryear{Chapman, Snowberg, Wang, and Camerer}{Chapman
  et~al.}{2018}]{chapman2018loss}
Chapman, J., E.~Snowberg, S.~Wang, and C.~Camerer (2018).
\newblock Loss attitudes in the us population: Evidence from dynamically
  optimized sequential experimentation (dose).
\newblock Technical report, National Bureau of Economic Research.

\bibitem[\protect\citeauthoryear{Christakis and Fowler}{Christakis and
  Fowler}{2007}]{christakis2007spread}
Christakis, N.~A. and J.~H. Fowler (2007).
\newblock The spread of obesity in a large social network over 32 years.
\newblock {\em New England Journal of Medicine\/}~{\em 357\/}(4), 370--379.

\bibitem[\protect\citeauthoryear{Christakis and Fowler}{Christakis and
  Fowler}{2008}]{christakis2008collective}
Christakis, N.~A. and J.~H. Fowler (2008).
\newblock The collective dynamics of smoking in a large social network.
\newblock {\em New England Journal of Medicine\/}~{\em 358\/}(21), 2249--2258.

\bibitem[\protect\citeauthoryear{Cohen, Frazzini, and Malloy}{Cohen
  et~al.}{2008}]{cohen2008small}
Cohen, L., A.~Frazzini, and C.~Malloy (2008).
\newblock The small world of investing: Board connections and mutual fund
  returns.
\newblock {\em Journal of Political Economy\/}~{\em 116\/}(5), 951--979.

\bibitem[\protect\citeauthoryear{Conti, Galeotti, Mueller, and Pudney}{Conti
  et~al.}{2013}]{conti2013popularity}
Conti, G., A.~Galeotti, G.~Mueller, and S.~Pudney (2013).
\newblock Popularity.
\newblock {\em Journal of Human Resources\/}~{\em 48\/}(4), 1072--1094.

\bibitem[\protect\citeauthoryear{Croson and Gneezy}{Croson and
  Gneezy}{2009}]{croson2009gender}
Croson, R. and U.~Gneezy (2009).
\newblock Gender differences in preferences.
\newblock {\em Journal of Economic Literature\/}~{\em 47\/}(2), 448.

\bibitem[\protect\citeauthoryear{Currarini, Jackson, and Pin}{Currarini
  et~al.}{2009}]{currarini2009economic}
Currarini, S., M.~O. Jackson, and P.~Pin (2009).
\newblock An economic model of friendship: Homophily, minorities, and
  segregation.
\newblock {\em Econometrica\/}~{\em 77\/}(4), 1003--1045.

\bibitem[\protect\citeauthoryear{Currarini, Jackson, and Pin}{Currarini
  et~al.}{2010}]{currarini2010identifying}
Currarini, S., M.~O. Jackson, and P.~Pin (2010).
\newblock Identifying the roles of race-based choice and chance in high school
  friendship network formation.
\newblock {\em Proceedings of the National Academy of Sciences\/}~{\em
  107\/}(11), 4857--4861.

\bibitem[\protect\citeauthoryear{De~Giorgi, Pellizzari, and Redaelli}{De~Giorgi
  et~al.}{2010}]{de2010identification}
De~Giorgi, G., M.~Pellizzari, and S.~Redaelli (2010).
\newblock Identification of social interactions through partially overlapping
  peer groups.
\newblock {\em American Economic Journal: Applied Economics\/}~{\em 2\/}(2),
  241--75.

\bibitem[\protect\citeauthoryear{De~la Haye, Robins, Mohr, and Wilson}{De~la
  Haye et~al.}{2010}]{de2010obesity}
De~la Haye, K., G.~Robins, P.~Mohr, and C.~Wilson (2010).
\newblock Obesity-related behaviors in adolescent friendship networks.
\newblock {\em Social Networks\/}~{\em 32\/}(3), 161--167.

\bibitem[\protect\citeauthoryear{Ductor, Goyal, and Prummer}{Ductor
  et~al.}{2018}]{ductor2018gender}
Ductor, L., S.~Goyal, and A.~Prummer (2018).
\newblock Gender \& collaboration.
\newblock Technical report, Working Paper.

\bibitem[\protect\citeauthoryear{Fafchamps, Goyal, and van~der Leij}{Fafchamps
  et~al.}{2010}]{fafchamps2010matching}
Fafchamps, M., S.~Goyal, and M.~J. van~der Leij (2010).
\newblock Matching and network effects.
\newblock {\em Journal of the European Economic Association\/}~{\em 8\/}(1),
  203--231.

\bibitem[\protect\citeauthoryear{Frederick}{Frederick}{2005}]{frederick2005cognitive}
Frederick, S. (2005).
\newblock Cognitive reflection and decision making.
\newblock {\em Journal of Economic perspectives\/}~{\em 19\/}(4), 25--42.

\bibitem[\protect\citeauthoryear{Girard, Hett, and Schunk}{Girard
  et~al.}{2015}]{girard2015}
Girard, Y., F.~Hett, and D.~Schunk (2015).
\newblock How individual characteristics shape the structure of social
  networks.
\newblock {\em Journal of Economic Behavior \& Organization\/}~{\em 115},
  197--216.

\bibitem[\protect\citeauthoryear{Glaeser, Sacerdote, et~al.}{Glaeser
  et~al.}{2000}]{glaeser2000social}
Glaeser, E.~L., B.~Sacerdote, et~al. (2000).
\newblock The social consequences of housing.
\newblock {\em Journal of Housing Economics\/}~{\em 9\/}(1-2), 1--23.

\bibitem[\protect\citeauthoryear{Granovetter}{Granovetter}{1973}]{granovetter1973strength}
Granovetter, M. (1973).
\newblock {{The Strength of Weak Ties}}.
\newblock {\em American Journal of Sociology\/}, 1360--1380.

\bibitem[\protect\citeauthoryear{Guterbock and London}{Guterbock and
  London}{1983}]{guterbock1983race}
Guterbock, T.~M. and B.~London (1983).
\newblock Race, political orientation, and participation: An empirical test of
  four competing theories.
\newblock {\em American Sociological Review\/}, 439--453.

\bibitem[\protect\citeauthoryear{Hoeting, Madigan, Raftery, and
  Volinsky}{Hoeting et~al.}{1999}]{hoeting1999bayesian}
Hoeting, J.~A., D.~Madigan, A.~E. Raftery, and C.~T. Volinsky (1999).
\newblock Bayesian model averaging: a tutorial.
\newblock {\em Statistical Science\/}, 382--401.

\bibitem[\protect\citeauthoryear{Hsieh and Lee}{Hsieh and
  Lee}{2016}]{hsieh2016social}
Hsieh, C.-S. and L.~F. Lee (2016).
\newblock A social interactions model with endogenous friendship formation and
  selectivity.
\newblock {\em Journal of Applied Econometrics\/}~{\em 31\/}(2), 301--319.

\bibitem[\protect\citeauthoryear{Hsieh and Van~Kippersluis}{Hsieh and
  Van~Kippersluis}{2018}]{hsieh2018}
Hsieh, C.-S. and H.~Van~Kippersluis (2018).
\newblock Smoking initiation: Peers and personality.
\newblock {\em Quantitative Economics\/}~{\em 9\/}(2), 825--863.

\bibitem[\protect\citeauthoryear{Ibarra}{Ibarra}{1992}]{ibarra1992homophily}
Ibarra, H. (1992).
\newblock Homophily and differential returns: Sex differences in network
  structure and access in an advertising firm.
\newblock {\em Administrative Science Quarterly\/}, 422--447.

\bibitem[\protect\citeauthoryear{Ibarra}{Ibarra}{1995}]{ibarra1995race}
Ibarra, H. (1995).
\newblock Race, opportunity, and diversity of social circles in managerial
  networks.
\newblock {\em Academy of Management\/}~{\em 38\/}(3), 673--703.

\bibitem[\protect\citeauthoryear{Jackson, Rogers, and Zenou}{Jackson
  et~al.}{2017}]{jackson2017economic}
Jackson, M.~O., B.~W. Rogers, and Y.~Zenou (2017).
\newblock The economic consequences of social-network structure.
\newblock {\em Journal of Economic Literature\/}~{\em 55\/}(1), 49--95.

\bibitem[\protect\citeauthoryear{Kossinets and Watts}{Kossinets and
  Watts}{2009}]{kossinets2009origins}
Kossinets, G. and D.~J. Watts (2009).
\newblock Origins of homophily in an evolving social network.
\newblock {\em American Journal of Sociology\/}~{\em 115\/}(2), 405--450.

\bibitem[\protect\citeauthoryear{Kov{\'a}{\v{r}}{\'\i}k, Bra{\~n}as-Garza,
  Cobo-Reyes, Espinosa, Jim{\'e}nez, and Ponti}{Kov{\'a}{\v{r}}{\'\i}k
  et~al.}{2012}]{kovavrik2012}
Kov{\'a}{\v{r}}{\'\i}k, J., P.~Bra{\~n}as-Garza, R.~Cobo-Reyes, M.~P. Espinosa,
  N.~Jim{\'e}nez, and G.~Ponti (2012).
\newblock Prosocial norms and degree heterogeneity in social networks.
\newblock {\em Physica A: Statistical mechanics and its Applications\/}~{\em
  391\/}(3), 849--853.

\bibitem[\protect\citeauthoryear{Kov{\'a}{\v{r}}{\'\i}k, Branas-Garza,
  Davidson, Haim, Carcelli, and Fowler}{Kov{\'a}{\v{r}}{\'\i}k
  et~al.}{2017}]{kovavrik2017}
Kov{\'a}{\v{r}}{\'\i}k, J., P.~Branas-Garza, M.~W. Davidson, D.~A. Haim,
  S.~Carcelli, and J.~H. Fowler (2017).
\newblock Digit ratio (2d: 4d) and social integration: an effect of prenatal
  sex hormones.
\newblock {\em Network Science\/}~{\em 5\/}(4), 476--489.

\bibitem[\protect\citeauthoryear{Kov{\'a}{\v{r}}{\'\i}k and Van~der
  Leij}{Kov{\'a}{\v{r}}{\'\i}k and Van~der Leij}{2014}]{kovarik2014risk}
Kov{\'a}{\v{r}}{\'\i}k, J. and M.~J. Van~der Leij (2014).
\newblock Risk aversion and social networks.
\newblock {\em Review of Network Economics\/}~{\em 13\/}(2), 121--155.

\bibitem[\protect\citeauthoryear{Kretschmer, Leszczensky, and Pink}{Kretschmer
  et~al.}{2018}]{kretschmer2018selection}
Kretschmer, D., L.~Leszczensky, and S.~Pink (2018).
\newblock Selection and influence processes in academic achievement—more
  pronounced for girls?
\newblock {\em Social Networks\/}~{\em 52}, 251--260.

\bibitem[\protect\citeauthoryear{Kruse, Smith, van Tubergen, and Maas}{Kruse
  et~al.}{2016}]{kruse2016neighbors}
Kruse, H., S.~Smith, F.~van Tubergen, and I.~Maas (2016).
\newblock From neighbors to school friends? how adolescents’ place of
  residence relates to same-ethnic school friendships.
\newblock {\em Social Networks\/}~{\em 44}, 130--142.

\bibitem[\protect\citeauthoryear{Lazarsfeld, Merton, et~al.}{Lazarsfeld
  et~al.}{1954}]{lazarsfeld1954friendship}
Lazarsfeld, P.~F., R.~K. Merton, et~al. (1954).
\newblock Friendship as a social process: A substantive and methodological
  analysis.
\newblock {\em Freedom and Control in Modern Society\/}~{\em 18\/}(1), 18--66.

\bibitem[\protect\citeauthoryear{Ley and Steel}{Ley and
  Steel}{2009}]{ley2009effect}
Ley, E. and M.~F. Steel (2009).
\newblock On the effect of prior assumptions in bayesian model averaging with
  applications to growth regression.
\newblock {\em Journal of Applied Econometrics\/}, 651--674.

\bibitem[\protect\citeauthoryear{Lindenlaub and Prummer}{Lindenlaub and
  Prummer}{2019}]{lindenlaub2019}
Lindenlaub, I. and A.~Prummer (2019).
\newblock Network structure and performance.
\newblock {\em The Economic Journal\/}.

\bibitem[\protect\citeauthoryear{Lorant and Tranmer}{Lorant and
  Tranmer}{2019}]{lorant2019peer}
Lorant, V. and M.~Tranmer (2019).
\newblock Peer, school, and country variations in adolescents’ health
  behaviour: A multilevel analysis of binary response variables in six european
  cities.
\newblock {\em Social Networks\/}~{\em 59}, 31--40.

\bibitem[\protect\citeauthoryear{Magnus, Powell, and Pr{\"u}fer}{Magnus
  et~al.}{2010}]{magnus2010comparison}
Magnus, J.~R., O.~Powell, and P.~Pr{\"u}fer (2010).
\newblock A comparison of two model averaging techniques with an application to
  growth empirics.
\newblock {\em Journal of Econometrics\/}~{\em 154\/}(2), 139--153.

\bibitem[\protect\citeauthoryear{Manski}{Manski}{1993}]{manski1993identification}
Manski, C.~F. (1993).
\newblock Identification of endogenous social effects: The reflection problem.
\newblock {\em The Review of Economic Studies\/}~{\em 60\/}(3), 531--542.

\bibitem[\protect\citeauthoryear{Marmaros and Sacerdote}{Marmaros and
  Sacerdote}{2002}]{marmaros2002peer}
Marmaros, D. and B.~Sacerdote (2002).
\newblock {{Peer and Social Networks in Job Search}}.
\newblock {\em European Economic Review\/}~{\em 46\/}(4), 870--879.

\bibitem[\protect\citeauthoryear{Marmaros and Sacerdote}{Marmaros and
  Sacerdote}{2006}]{marmaros2006friendships}
Marmaros, D. and B.~Sacerdote (2006).
\newblock How do friendships form?
\newblock {\em The Quarterly Journal of Economics\/}~{\em 121\/}(1), 79--119.

\bibitem[\protect\citeauthoryear{Mayer and Puller}{Mayer and
  Puller}{2008}]{mayer2008old}
Mayer, A. and S.~L. Puller (2008).
\newblock The old boy (and girl) network: Social network formation on
  university campuses.
\newblock {\em Journal of Public Economics\/}~{\em 92\/}(1-2), 329--347.

\bibitem[\protect\citeauthoryear{McPherson, Smith-Lovin, and Cook}{McPherson
  et~al.}{2001}]{mcpherson2001birds}
McPherson, M., L.~Smith-Lovin, and J.~M. Cook (2001).
\newblock Birds of a feather: Homophily in social networks.
\newblock {\em Annual Review of Sociology\/}~{\em 27\/}(1), 415--444.

\bibitem[\protect\citeauthoryear{Moody}{Moody}{2001}]{moody2001race}
Moody, J. (2001).
\newblock Race, school integration, and friendship segregation in america.
\newblock {\em American Journal of Sociology\/}~{\em 107\/}(3), 679--716.

\bibitem[\protect\citeauthoryear{Patacchini and Zenou}{Patacchini and
  Zenou}{2016}]{patacchini2016racial}
Patacchini, E. and Y.~Zenou (2016).
\newblock Racial identity and education in social networks.
\newblock {\em Social Networks\/}~{\em 44}, 85--94.

\bibitem[\protect\citeauthoryear{Preciado, Snijders, Burk, Stattin, and
  Kerr}{Preciado et~al.}{2012}]{preciado2012does}
Preciado, P., T.~A. Snijders, W.~J. Burk, H.~Stattin, and M.~Kerr (2012).
\newblock Does proximity matter? distance dependence of adolescent friendships.
\newblock {\em Social networks\/}~{\em 34\/}(1), 18--31.

\bibitem[\protect\citeauthoryear{Ruef, Aldrich, and Carter}{Ruef
  et~al.}{2003}]{ruef2003structure}
Ruef, M., H.~E. Aldrich, and N.~M. Carter (2003).
\newblock The structure of founding teams: Homophily, strong ties, and
  isolation among us entrepreneurs.
\newblock {\em American Sociological Review\/}, 195--222.

\bibitem[\protect\citeauthoryear{Sala-i Martin, Doppelhofer, and Miller}{Sala-i
  Martin et~al.}{2004a}]{sala2004}
Sala-i Martin, X., G.~Doppelhofer, and R.~I. Miller (2004a).
\newblock Determinants of long-term growth: A bayesian averaging of classical
  estimates (bace) approach.
\newblock {\em The American Economic Review\/}, 813--835.

\bibitem[\protect\citeauthoryear{Sala-i Martin, Doppelhofer, and Miller}{Sala-i
  Martin et~al.}{2004b}]{sala2004determinants}
Sala-i Martin, X., G.~Doppelhofer, and R.~I. Miller (2004b).
\newblock Determinants of long-term growth: A bayesian averaging of classical
  estimates (bace) approach.
\newblock {\em The American Economic Review\/}, 813--835.

\bibitem[\protect\citeauthoryear{Sharot}{Sharot}{2011}]{sharot2011optimism}
Sharot, T. (2011).
\newblock The optimism bias.
\newblock {\em Current Biology\/}~{\em 21\/}(23), R941--R945.

\bibitem[\protect\citeauthoryear{Simmons, Nelson, and Simonsohn}{Simmons
  et~al.}{2011}]{simmons2011}
Simmons, J.~P., L.~D. Nelson, and U.~Simonsohn (2011).
\newblock False-positive psychology: Undisclosed flexibility in data collection
  and analysis allows presenting anything as significant.
\newblock {\em Psychological Science\/}~{\em 22\/}(11), 1359--1366.

\bibitem[\protect\citeauthoryear{Valente, Fujimoto, Chou, and
  Spruijt-Metz}{Valente et~al.}{2009}]{valente2009adolescent}
Valente, T.~W., K.~Fujimoto, C.-P. Chou, and D.~Spruijt-Metz (2009).
\newblock Adolescent affiliations and adiposity: a social network analysis of
  friendships and obesity.
\newblock {\em Journal of Adolescent Health\/}~{\em 45\/}(2), 202--204.

\bibitem[\protect\citeauthoryear{Wittek, Kroneberg, and L{\"a}mmermann}{Wittek
  et~al.}{2020}]{wittek2020fighting}
Wittek, M., C.~Kroneberg, and K.~L{\"a}mmermann (2020).
\newblock Who is fighting with whom? how ethnic origin shapes friendship,
  dislike, and physical violence relations in german secondary schools.
\newblock {\em Social Networks\/}~{\em 60}, 34--47.

\bibitem[\protect\citeauthoryear{Zeltzer}{Zeltzer}{2020}]{zeltzer2019gender}
Zeltzer, D. (2020).
\newblock Gender homophily in referral networks: Consequences for the medicare
  physician earnings gap.
\newblock {\em American Economic Journal: Applied Economics\/}, forthcoming.

\end{thebibliography}
}

\section*{Appendix}

\begin{table}[htp!]
\caption{Determinants of link formation in students' network: female sample}
\centering
\vspace{0.5cm}
\begin{tabular}[c]{lccc}\hline
\\
& PI prob. & Pt. Mean & Pt. Std. \\ \hline
\\
\textbf{Common Gender}&1& 0.025 & 0.004\\[4pt]
\textbf{Common Section}&1& 0.061 &0.005\\[4pt]
\textbf{Friends$_{t-1}$} &1&0.395 &0.036\\[4pt]  
\textbf{Both Smokers}&0.99 &0.049&0.012\\[4pt]
Altruism diff.&0.22&-0.001&0.003\\[4pt]
Inconsistent diff. & 0.22 & -0.003& 0.006\\[4pt]
Parent educ. diff.&0.09& -0.0003 &0.0011\\[4pt]
CRT diff. & 0.07 & -0.0004 & 0.0016\\[4pt]
Both Reflective &0.09&0.0011&0.0037\\[4pt]
Time pref. diff. &0.03&-0.0000&0.0003\\[4pt]
Reciprocity diff.&0.05& 0.0000 &0.0005\\[4pt]
Both altruism learner&0.04& -0.0002 &0.0013\\[4pt]
Both STEM best grade&0.03& 0.0002 &0.0017\\[4pt]
Risk diff.&0.02& -0.0000 &0.0002\\[4pt]
Income diff. &0.02&-0.0000&0.0003\\[4pt]
Self-confidence diff.&0.02& 0.0000 &0.0003\\[4pt]
BMI diff.&0.02& 0.0000 &0.0001\\[4pt]
Both volunteers&0.02& 0.0000 &0.0021\\[4pt]
Both STEM pref. &0.02& 0.0000 &0.0010\\[4pt]
Both right&0.02& -0.0000 &0.0009\\[4pt]
Observations & 4092 & 4092 & 4092\\
\hline
\\
\end{tabular}
\begin{minipage}{14cm}
   \footnotesize  \textit{Notes}: Column 1 presents the posterior inclusion probability ($PIP$). Robust determinants are those with $PIP \geq 0.8$ in bold. Column 2 shows the posterior mean. Column 3 reports the posterior standard deviation. The dependent variable is equal to 1 if $i$ and $j$ are friends in Period 2, 0 otherwise. The results are obtained by using a uniform prior for the prior model probability and a BRIC prior for the hyperparameter that measures the degree of prior uncertainty on coefficients, $g=1/max(N,K^{2})$.
  \end{minipage}
\label{Table BMAfemale}
\end{table}

\begin{table}[htp!]
\caption{Determinants of link formation in students' network: Male sample}
\centering
\vspace{0.5cm}
\begin{tabular}[c]{lccc}\hline
\\
& PI prob. & Pt. Mean & Pt. Std. \\ \hline
\\
\textbf{Common Section}&1& 0.068 &0.005\\[4pt]
\textbf{Friends$_{t-1}$} &1&0.457 &0.030\\[4pt]  
Both right& 0.08& 0.0008 &0.003\\[4pt]
Common Gender & 0.07 & 0.005 & 0.002\\[4pt]
Both altruism learner&0.05& 0.0003 &0.0017\\[4pt]
Parent educ. diff.&0.05&0.0001 &0.0006\\[4pt]
Self-confidence diff.&0.04& -0.0001 &0.0006\\[4pt]
BMI diff.&0.04& -0.0000 &0.0003\\[4pt]
Reciprocity diff.&0.02& -0.0000 &0.0004\\[4pt]
CRT diff. & 0.02 & -0.0000 & 0.0004\\[4pt]
Risk diff.&0.02& 0.0000 &0.0002\\[4pt]
Both Smokers &0.02&0.0002&0.0025\\[4pt]
Altruism diff.&0.02&-0.0000&0.0003\\[4pt]
Time pref. diff. &0.02&-0.0000&0.0001\\[4pt]
Income diff. &0.02&-0.0000&0.0003\\[4pt]
Both volunteers&0.02& -0.0001 &0.0017\\[4pt]
Inconsistent diff. & 0.02 & 0.0000& 0.0008\\[4pt]
Both Reflective &0.02&0.0000&0.0009\\[4pt]
Both STEM best grade&0.02& -0.0001 &0.0014\\[4pt]
Both STEM pref. &0.02& -0.0000 &0.0008\\[4pt]
Observations & 4423 & 4423 & 4423\\
\hline
\\
\end{tabular}
\begin{minipage}{14cm}
   \footnotesize  \textit{Notes}: Column 1 presents the posterior inclusion probability ($PIP$). Robust determinants are those with $PIP \geq 0.8$ in bold. Column 2 shows the posterior mean. Column 3 reports the posterior standard deviation. The dependent variable is equal to 1 if $i$ and $j$ are friends in Period 2, 0 otherwise. The results are obtained by using a uniform prior for the prior model probability and a BRIC prior for the hyperparameter that measures the degree of prior uncertainty on coefficients, $g=1/max(N,K^{2})$.
  \end{minipage}
\label{Table BMAmale}
\end{table}

\end{document}